\documentclass[prb,twocolumn,showpacs,nofootinbib]{revtex4}
\usepackage[dvips]{graphicx}
\usepackage{color,array,dcolumn}
\usepackage{amsmath}
\usepackage{amssymb}
\usepackage{amsmath}
\usepackage{amssymb}
\begin{document}
\title{Quantum Monte Carlo study of the two-dimensional electron gas in presence of Rashba interaction}
\author{A. Ambrosetti}
\email{ambrosetti@science.unitn.it}
\affiliation{Dipartimento di Fisica, University of Trento, via Sommarive 14, I--38050, 
Povo, Trento, Italy}
\affiliation{INFN, Gruppo Collegato di Trento, Trento, Italy}
\author{F.Pederiva}
\email{pederiva@science.unitn.it}
\affiliation{Dipartimento di Fisica, University of Trento, via Sommarive 14, I--38050, 
Povo, Trento, Italy}
\affiliation{INFN, Gruppo Collegato di Trento, Trento, Italy}
\author{E. Lipparini}
\email{lipparin@science.unitn.it}
\affiliation{Dipartimento di Fisica, University of Trento, via Sommarive 14, I--38050, 
Povo, Trento, Italy}
\affiliation{INFN, Gruppo Collegato di Trento, Trento, Italy}
\author{S. Gandolfi}
\email{gandolfi@sissa.it}
\affiliation{International School for Advanced Studies, SISSA
Via Beirut 2/4 I-34014 Trieste, Italy}
\affiliation{INFN, Sezione di Trieste, Trieste, Italy}
\begin{abstract}

\date{\today}

We introduce a variant to the Diffusion Monte Carlo algorithm that can be
employed to study the effects of the Rashba interaction in a many electron 
systems. Because of the
spin--orbit nature of Rashba interaction a standard algorithm cannot be
applied and therefore a specific imaginary time spin dependent
propagator has been developed and implemented following previous work
developed in the framework of nuclear physics. We computed the ground
state energy of the 2D electron gas at different densities for several
values of the Rashba interaction
strength as a function of ``Rashba spin states'' polarization. Comparison
is given with analytically known Hartree-Fock results and for the system
in absence of Coulomb interaction.

\end{abstract}

\maketitle

\section{Introduction}

The Rashba interaction, an electric field induced spin-orbit coupling,
has been experimentally observed in semiconductor hetero-structures,
depending on their symmetry, and has been proved to be tunable in strength
through a gate voltage \cite{nitta,engels}. The external voltage can then be 
used to control the spin state of the system. This property becomes extremely
interesting in view of spintronics applications. Several experiments have
also been performed with the aim of studying the dependence of the interaction
strength on applied gate voltage and well thickness \cite{nittath}.

The typical experimental setup consists of a device etched on
a two-dimensional quantum well formed at the interface of two
semiconductors. In such devices electrons are constrained to move
in a two-dimensional space ($O\hat{x}-O\hat{y}$) forming a 2D gas. The
asymmetry of the quantum well generates an electric field, along the 
$\hat z$--direction, perpendicular
to the plane containing the electrons. This causes electrons to be subject to
an effective magnetic field $\mathbf{B}_{eff}\propto\mathbf{p}\times\mathbf{E}$
coupling to their spin. Such coupling gives rise to the well known Rashba potential \cite{Rashba}: 
\begin{equation}
V_{Rashba}=\lambda \sum_{i=1}^N [p_i^{y}\sigma_i^{x}-p_i^{x}\sigma_i^{y}] \,, 
\label{vrashba}
\end{equation} 
where $\mathbf p_i$ is the momentum of the $i-th$ electron, 
and $\sigma_i^x$ and $\sigma_i^y$ are the Pauli matrices acting over 
the spin of particle $i$. Neglecting the Coulomb interaction among the electrons the Hamiltonian 
is a sum of one body terms, and the problem is analytically solvable. 
Single particle solutions are given by plane waves with $\mathbf k$--dependent
spinors. In particular, for each wave vector with momentum 
$\mathbf k$ two possible solutions exist:
\begin{equation}
\chi (\mathbf r) = \frac{e^{i\mathbf k\cdot \mathbf r}}{\sqrt{2}}
\left (
\begin{array}{c}
\pm \frac{k_y+ik_x}{k} \\
1 \\
\end{array}
\right ) \,.
\label{eq:rashbawf}
\end{equation}
These solutions correspond to two different spin states,
with the following dispersion law:
\begin{equation}
\varepsilon (k) = \frac{\mathbf{k}^2}{2m}\pm \lambda |\mathbf{k}| \,.
\end{equation}

This spin splitting causes non interacting electrons to arrange in two bands with
different fillings depending on the interaction strength $\lambda$, therefore inducing
a natural polarization of the system. In this case two different
Fermi surfaces are generated, as also shown experimentally by beating
patterns in the Shubnikov-De Haas oscillations \cite{nitta,engels}.
Being $\lambda$ proportional to the intensity of the external electric field,
it is possible to manipulate the occupation numbers and to induce at the
same time precession on the electron spin. These properties induced new
studies in the direction of spin field effect transistors\cite{Datta}.

Once the Coulomb interaction is included in the Hamiltonian, no
analytic solution is available. Several methods can be used to study
this system and much theoretical work has been done so far (for a review see e.g. \cite{lip08}).
The Hartree-Fock method gives a simple analytical
solution, and though it totally ignores the effect of correlations,  it provides
a useful insight on the structure of the single particle levels.
Another interesting approach consists
in applying a unitary transformation $U$ \cite{Aleiner,Valin} giving, to leading order in
the spin-orbit strength, a transformed Hamiltonian $\tilde{H}=U^{-1}HU$
whose eigenstates are also spin and angular momentum eigenstates. Though
approximate, this method allows the use of standard Quantum Monte Carlo
techniques\cite{transform}.

In this work a particular implementation of the Diffusion Monte Carlo,
directly dealing with spin--orbit interactions, is proposed in
order to obtain an ab--initio method giving accurate ground state energies
of the system. No transformation is needed because the spin-orbit term
is included in the imaginary time propagator.\\

The paper is organized as
follows: in Sec. \ref{secmeth} we describe the Diffusion Monte Carlo
method and the way in which the Green's function used for treating Rashba 
interaction is derived. The trial wave function used as a starting point for the
projection is described in Sec. \ref{wavefunction}. Finally in Sec. 
\ref{secres} the results are presented and discussed together with Hartree-Fock energies 
for the ground state of the system, and conclusions are given in Sec. \ref{conclusions}.

\section{Method} 
\label{secmeth}

\subsection{Diffusion Monte Carlo method for general spin--independent 
many--body problems}

The Diffusion Monte Carlo (DMC) method is based on imaginary time
evolution projection. 
Through the use of an appropriate Green's function,
an initial state is projected over its lowest energy component having
the same nodes or the same phase as the trial wave function $\psi_T$. 
The propagation in imaginary time $\tau$ is given by
\begin{eqnarray}
\label{eq:a}
\psi(\mathbf R,\tau)&=&\exp{\left[-(H-E_0)\tau\right]}\psi(\mathbf R,0)=
\nonumber \\
&&\sum_n c_n \exp{\left[-(H-E_0)\tau\right]}\phi_n(\mathbf R) \,,
\end{eqnarray}
where $E_0$ is a normalization factor, and $\mathbf R$ represents the
space coordinates of the system.
The amplitudes of higher energy
states, due to propagation, decay exponentially with $\tau$, with a
lifetime inversely proportional to their energy relative to $E_0$.
If $\psi(\mathbf R ,0)$ is not orthogonal to the ground state, 
$\psi(\mathbf R ,\tau \rightarrow \infty)$ will be proportional to the ground state itself.

A typical many--body system is described with a Hamiltonian of the form
\begin{equation}
H=\hat{T} + \hat{V} \,,
\end{equation}
where $\hat{T}$ is the sum of single--particle kinetic energy operators, and 
the potential $\hat{V}$ between all the electrons only depends on spatial 
coordinates. The 
evolution in imaginary time can be achieved by means of the Green's Function 
of the Hamiltonian that can be approximated by using Trotter's 
formula:
\begin{equation}
 e^{-\tau \hat{H}} = e^{-\tau \hat{V}} e^{-\tau \hat{T}} + O(\tau) \,.
\label{trott}
\end{equation}
In space representation the propagator \eqref{trott} can be written as:
\begin{eqnarray}
 G(\mathbf{R},\mathbf{R'},\tau) = \langle \mathbf{R}| e^{-(\hat{H}-E_0)
 \tau}|\mathbf{R'}\rangle 
\nonumber \\
 \simeq e^{[V(\mathbf{R})-E_0]\tau} G_0
 (\mathbf{R},\mathbf{R'},\tau) \,.
\end{eqnarray}
$G_0$ is the exact Green's function of a two--dimensional 
non--interacting system:
\begin{equation}
 G_0 (\mathbf{R},\mathbf{R'},\tau) = \frac{1}{4 \pi D \tau} e^{ \big[
 \frac{(\mathbf{R}-\mathbf{R'})^2}{2 D\tau} \big]} \,,
\end{equation}
where $D=\hbar^2/m$ is the diffusion constant.
The exponential of the kinetic term gives rise to a free particle
imaginary time Green's function $G_0$, while the other term is viewed
as a weighting term.

The algorithm makes use of walkers, i.e. points in the coordinate space,
in order to sample the ground state wave function. The propagation is
obtained by diffusing the walkers according the displacements distribution
given by $G_0$. Afterwards a weight is assigned to the walkers according
to the other factor in the propagator given by the potential energy of the
system. 
Because of the Trotter approximation, the Green's function is
correct only at order $O(\tau)$. The problem is overcome by using short
time propagation repeatedly in order to achieve long enough imaginary
times and adequate statistics, minimizing at the same time the time step
error.
The result is given after an extrapolation to $\Delta\tau\rightarrow0$.

In order to enhance the efficiency, importance sampling is
implemented in DMC algorithms \cite{Grimm,cep79}. The idea consists in sampling a density
of points proportional to the ground state distribution multiplied by
an importance function. In standard cases the importance function will
only depend on space coordinates being the spin fixed, thus:
\begin{eqnarray}
&&\psi_T (\mathbf{R}) \phi (\mathbf{R},\tau)=
\nonumber \\
&&\int G(\mathbf{R},\mathbf{R'},\tau) 
\frac{\psi_T(\mathbf{R})}{\psi_T(\mathbf{R'})} \psi_T(\mathbf{R'})
\phi(\mathbf{R'},0) d\mathbf{R}\,.
\end{eqnarray}
This can be shown to introduce a drift in $G_0$ and to modify the
weighting factor, which will not only contain the potential, but also the 
local energy $E_L=H\psi_T(\mathbf{R})/\psi_T(\mathbf{R})$ of the system. 

\subsection{Many--body spin--dependent Hamiltonian}
The system we are studying is a two-dimensional electron gas at $T=0$, in
presence of both Coulomb and Rashba interaction, with the addition of a
uniform charge background \cite{ceperley,tanatar}. 
The Hamiltonian for the system can be written as
\begin{equation}
H=\sum^{N}_{i=1} \frac{P^2 _i}{2m}+\lambda \sum^{N}_{i=1}(p_i^y
\sigma _i^x - p_i^x \sigma _i^y)+ V_{Coul}(\mathbf{R}) \,,
\label{hamiltoniannors} 
\end{equation}
where $V_{Coul}$ includes the electron--electron interaction and the effects
of the background.
The Hamiltonian in this case does not only contain space coordinate
dependent potentials. The Rashba interaction, because of its spin-orbit
character, contains spin and momentum operators, which therefore cannot
be treated like simple weighting factors as in the case shown above. In
order to apply the DMC technique, a new propagator form is needed, 
taking into account the particular features of non locality and spin-dependence 
\cite{sarsa03,gandolfi09}. 

In order to simplify the following calculations let us first consider 
the single-particle Hamiltonian given by the 
kinetic term and the Rashba interaction for only one electron. 
By applying Trotter's formula
the following form of the Green's function can be obtained
\begin{equation}
G(\mathbf{r},\mathbf{r'},\Delta \tau)= e^{-\lambda (p_y
\sigma _x -p_x \sigma _y) \Delta \tau} G_0(\mathbf{r},\mathbf{r'},\Delta
\tau) \,,
\label{eq:rashbaprop}
\end{equation}
where the Pauli matrices $\sigma_x$ and $\sigma_y$ act on the spin 
components of the electron, and $\mathbf{r}$ and $\mathbf{r'}$ are the coordinates
of the electron after and before the diffusion generated by $G_0$.
The last factor $G_0$ will again give the space displacement, while the second one,
including the Rashba interaction, contains both momentum and spin operators, and 
can be viewed as acting on the free propagator.
Expanding to first order in $\Delta \tau$ the first factor of Eq. 
\eqref{eq:rashbaprop}, and applying the derivatives given by 
$\mathbf p=-i\hbar\mathbf\nabla$ to $G_0$, we have
\begin{align}
\left[ 1 -i\frac{\lambda}{D} (\sigma_{x}\Delta y-\sigma_{y}\Delta x)\right]
G_0(\mathbf{r},\mathbf{r'},\Delta \tau) \simeq 
\nonumber \\
\exp{\left[-i\frac{\lambda}{D}(\sigma_{x}\Delta y -\sigma_{y} \Delta x)\right]} 
G_0(\mathbf{r},\mathbf{r'},\Delta \tau) \,.
\end{align}
The first part can be interpreted as a spin rotation depending on $\Delta
x=x-x'$ and $\Delta y=y-y'$, i.e. the displacement generated by the
Gaussian free particle Green's function $G_0$. 
The appearance of the spin-rotating factor in the Green's function implies
that the spin coordinates of the electrons must be explicitly used.
Making use of a spinorial representation, the spin-state for the $i-th$ electron is given by
\begin{equation}
\vert s_i\rangle=\alpha_i\lvert\uparrow\rangle+\beta_i\lvert\downarrow\rangle \,,
\end{equation}
where $\alpha$ and $\beta$ are the amplitudes of the spin state
in the $\{\lvert\uparrow\rangle\}$ and $\{\lvert\downarrow\rangle\}$ basis.

The propagation of the spin--dependent Green's function can then be realized 
with a rotation of spinors. For each electron $i$ the spin--dependent propagator 
can be written using the following matrix form:
\begin{eqnarray}
&&\exp{\left[-i\frac{\lambda}{D}(\sigma_{x}\Delta y -\sigma_{y} \Delta x)
\right]}\equiv O_i =
\nonumber \\
&&\left( 
\begin{array}{cc}
 \cos\left(\frac{\lambda}{D}\Delta r_i\right) & 
 \sin\left(\frac{\lambda}{D}\Delta r_i\right)\frac{-i\Delta y_i +\Delta x_i}{\Delta r_i}  \\
-\sin\left(\frac{\lambda}{D}\Delta r_i\right)\frac{i\Delta y_i +\Delta x_i}{\Delta r_i} & 
 \cos\left(\frac{\lambda}{D}\Delta r_i\right)  \\
\end{array} \right) \,,
\nonumber \\
\end{eqnarray}
where $\Delta r_i=\sqrt{\Delta x_i^2+\Delta y_i^2}$.
Therefore, during the propagation in imaginary time, an electron with initial coordinates
$(\mathbf{r_i},\alpha_i,\beta_i)$, will be first moved to $(\mathbf{r'_i},\alpha
_i,\beta _i)$ due to the free propagator $G_0$, and then to $(\mathbf{r'_i},\alpha
'_i,\beta '_i)$ due to rotation according to
\begin{equation}
(\alpha '_i\,,\beta '_i)={O_i}\left( 
\begin{array}{c} 
\alpha _i\\
\beta _i\\
\end{array} \right) \,.
\end{equation} 

The approximations introduced after Eq. \eqref{eq:rashbaprop} generate an error which must be now
taken into account at least to order $\Delta \tau$. 
If the spin-orbit propagator just derived were correct, we would expect to obtain
\begin{align}
\int e^{-i\frac{\lambda}{D}(\sigma_{x}\Delta y -\sigma_{y} \Delta x)} G_0(\mathbf{r},\mathbf{r'},\Delta \tau) 
\psi (\mathbf{r'}) d\mathbf{r'} = \nonumber \\
e^{-\lambda (p_y \sigma_x -p_x \sigma_y)} \int G_0(\mathbf{r},\mathbf{r'},\Delta \tau) \psi (\mathbf{r'}) d\mathbf{r'} \,.
\end{align}
This equation does not hold unless the propagator is corrected for a weighting factor
\begin{equation}
 e^{\frac{\lambda^2}{D}\Delta \tau}. 
\end{equation}
It is important to note that the results obtained following this
approach coincide with the analytical form of the Green's function
for the system in absence of Coulomb potential. In this case no error
is introduced in $G(\mathbf{R},\mathbf{R'}, \Delta \tau)$ because the Rashba
interaction commutes with the kinetic term of the Hamiltonian. 

By considering the full many--particle Hamiltonian, including the Coulomb interaction, 
the total Green's function is
\begin{align}
G(\mathbf{R},\mathbf{R'},\Delta\tau)=
e^{-\left[V_{Coul} (\mathbf{R})-E_0-\frac{N\lambda^2}{D}\right] \Delta \tau}\times
\nonumber \\
e^{-i\frac{\lambda}{D} \sum^N_{i=1}(\Delta r_i^y\sigma_i^x -\Delta r_i^x\sigma_i^y)\Delta\tau} 
G_0(\mathbf{R},\mathbf{R'},\Delta \tau)\,.
\end{align}

When a spin-orbit term is introduced in the Hamiltonian, as shown before,
the propagator will not leave spin unchanged because space coordinate
diffusion becomes related to spin rotation. This means that also the
importance sampling should not be naively applied, but in the total
weight, a factor $\psi_T(\mathbf{R},S)/\psi_T(\mathbf{R'},S')$, where $S'$
and $S$ respectively are the old and new spin states, should be taken
into account.
It is possible to apply the importance sampling adding a drift 
term in $G_0$ as usual by considering 
\begin{eqnarray}
&&\frac{\psi_T(\mathbf{R},S)}{\psi_T(\mathbf{R'},S')}=
\nonumber \\
&&\frac{\psi_T(\mathbf{R},S')}{\psi_T(\mathbf{R'},S')}
\frac{\psi_T(\mathbf{R},S)}{\psi_T(\mathbf{R},S')}
\approx
\frac{\psi_T(\mathbf{R},S')}{\psi_T(\mathbf{R'},S')}
\frac{\psi_T(\mathbf{R'},S)}{\psi_T(\mathbf{R'},S')} \,,
\label{eq:is}
\end{eqnarray}
where the two forms of Eq. \eqref{eq:is} are equivalent to first order in $\Delta\tau$.
The factor where spins are unchanged can be included as usual, while the second one
can be interpreted as an additional weighting factor. 
We point out that in this case the local energy appearing in the weight 
includes only the the spin--independent part of the Hamiltonian.

\subsection{Wavefunction}
\label{wavefunction}
We used a trial wave function made of a Slater
determinant of single particle states multiplied by a Jastrow factor
accounting for correlations, according to the form 
\begin{equation}
\psi_T(\mathbf{R},S)=D(\mathbf{R},S)\exp\left(-\sum^N_{i<j}
u(|\mathbf{r_i}-\mathbf{r_j}|)\right) \,,
\end{equation} 
where $u(r)$ is a two body pseudopotential of the double 
Yukawa form \cite{ceperley,tanatar}. 

Single particle states used to build the determinant are chosen
as the eigenstates of the Hamiltonian in absence of Coulomb interaction,
coinciding also with Hartree-Fock single particle solutions,
written in Eq. \eqref{eq:rashbawf}
\begin{equation}
\phi_{\mathbf{k},\pm}(\mathbf r,\alpha,\beta)= \left[\pm\alpha 
\frac{k_y+ik_x}{|\mathbf{k}|} + \beta
\right]e^{i \mathbf{k}\cdot \mathbf{r}} \,,
\label{frashba}
\end{equation}
where $\alpha$ and $\beta$ are respectively the up and down spin
components with respect to the $z$ axis. These are pairs of wave functions
which we will call \emph{quasi--up} (plus sign) and \emph{quasi--down} 
(minus sign) states. 

DMC simulations of infinite systems are usually performed in close
shell configurations in order to obtain a real wave function and to 
reduce the finite size effects mainly due to the kinetic energy. This makes
it possible to apply the fixed node approximation in order to obtain
a state with the same symmetry as $\psi_T$. In our case the trial
wave function cannot be reduced to a real form because of the phase
change induced by spin rotation. Therefore we employ the fixed--phase
approximation \cite{bolton,colletti,gandolfi07,gandolfi07b,gandolfi08b,gandolfi09}. 
For any complex wave function it is always possible to factorize the 
modulus in the following way
\begin{equation}
\psi_T(\mathbf{R},S)=|\psi_T(\mathbf{R},S)| e^{i\theta_T(\mathbf{R},S)} \,.
\end{equation}
Using fixed node approximation means finding
the lowest energy state $\phi(\mathbf{R})$ whose product with
the (in this case real) trial wave function is positively defined
$\phi(\mathbf{R})\psi^*_T(\mathbf{R})=\phi(\mathbf{R})\psi_T(\mathbf{R})>0$. 
In our case this product does not have a defined sign and cannot be used as
a sampling density: 
\begin{eqnarray}
&&\phi(\mathbf{R},S)\psi^*_T(\mathbf{R},S)=
\nonumber\\
&&|\phi(\mathbf{R},S)||\psi_T(\mathbf{R},S)|
\exp[i(\theta(\mathbf{R},S)-\theta_T(\mathbf{R},S))] \,.
\label{pphase}
\end{eqnarray}
In the fixed phase approximation the problem is overcome assuming 
\begin{equation}
\exp[i(\theta(\mathbf{R},S)-\theta_T(\mathbf{R},S))]=1 \,,
\end{equation}
which corresponds to finding a solution with the same
phase as the trial wave function. Under this assumption Eq. \eqref{pphase}
becomes positive definite and the sampled distribution will only depend
on the wave functions modulus. 

In order to calculate the ground state
energy at different polarizations and to reduce finite size effects,
Twist Averaged Boundary Conditions (TABC) have been introduced in
the algorithm \cite{twist,gandolfi09}. As shown in the electron gas without
spin--orbit 
interactions, this allows to obtain good results without the need to use
a very large number of electrons, giving therefore a great improvement
in terms of the computational time required.

\section{Results} \label{secres}

In order to verify our method a check has been performed by
comparing numerical results with the exact solution obtained neglecting
the electron--electron interaction. As mentioned in the
introduction, the Hamiltonian 
\begin{equation}
H_0=\sum^N_{i=1}\frac{P^2_i}{2m}+\lambda 
\sum_{i=1}^N [p^{y}_i\sigma_i^{x}-p^{x}_i\sigma_i^{y}] 
\label{hnorashba}
\end{equation}
can be diagonalized by the single particle wave functions
\eqref{frashba}. The analytical solution to $N$ fermions only subject to
Rashba interaction is therefore a $N\times N$ Slater determinant of these
single particle solutions. For a given $\lambda$ value, the ground state
has a well defined polarization. As a test for our DMC we multiplied
the exact wave function by a Slater determinant in order to obtain a
wrong state with the same nodes as the real ground state. Our algorithm
indeed proved to be able to project out the ground state obtaining the
analytical result within errorbars. 


In this work we make use of the dimensionless parameter $r_s=r_0/a_0$,
defined as a function of the Bohr radius $a_0=\hbar^2 /me^2$ and the
radius containing only one particle on average $V/N=\pi r_0 ^2$. Energies
are given in units of Rydberg/electron (1 $Ry = e^2 /2 a_0$) and the
Hamiltonian of Eq. \eqref{hamiltoniannors} in these units is
\begin{align}
H=\sum^{N}_{i=1} \left [ \frac{-\nabla ^2 _i}{r_s^2}-\frac{2 i\lambda}{r_s} 
(\partial_i^y \sigma _i^x - \partial_i^x \sigma _i^y)\right]+ \\
 \frac{2 e^2}{r_s} \sum^{N}_{i<j} \frac{1}{|\mathbf{r}_i-\mathbf{r}_j|}+V_{back}\,,
\label{hamiltonian}
\end{align}
where $V_{back}$ contains the effects of the charge background.

The main numerical results presented in this work concern the ground state
energies of the two--dimensional electron gas with Coulomb and Rashba interaction
at different densities (expressed in terms of the $r_s$ parameter),
different values of the intensity $\lambda$ of the Rashba interaction and
different polarizations defined as
\begin{equation}
\xi=\frac{N_+ -N_-}{N} \,,
\label{eq:pol}
\end{equation}
where $N_-$ and $N_+$ are the numbers of electrons in the quasi--down and quasi--up
spin states respectively, and the total number of electrons in the simulation is
fixed to $N=58$.
All the values obtained for the ground state energies as a function of $r_s$,
$\lambda$ and $\xi$ are reported in the appendix, where the DMC and the Hartree--Fock
results are compared.

In particular the $r_s$ values $1$, $5$, $10$ and $20$ were
chosen and, for each density, calculations were performed at different
$\lambda$ values of ground state energy as a function of quasi-up -
quasi-down occupation numbers, the total number of electrons in
our simulation being fixed to $N=58$. Coulomb interaction was treated by means
of the Ewald sums \cite{ewald}. As a comparison, besides DMC results,
we also give Hartree-Fock energies, which are given, in this case,
by the following analytic form for infinite systems:
\begin{align}
\frac{E}{N}=&\frac{(1-\xi^2)}{r^2_s}+\frac{2\sqrt{2}}{r^3_s}\Bigg[
(\lambda-\frac{2}{\pi})(1+\xi)^{3/2} 
\nonumber\\
&-(\lambda+\frac{2}{\pi})(1-\xi)^{3/2}\Bigg] \,.
\end{align}
where $\xi$ was defined in \eqref{eq:pol}. \\

\begin{figure}[ht]
\vspace{0.7cm}
\centering
\includegraphics[scale=0.32]{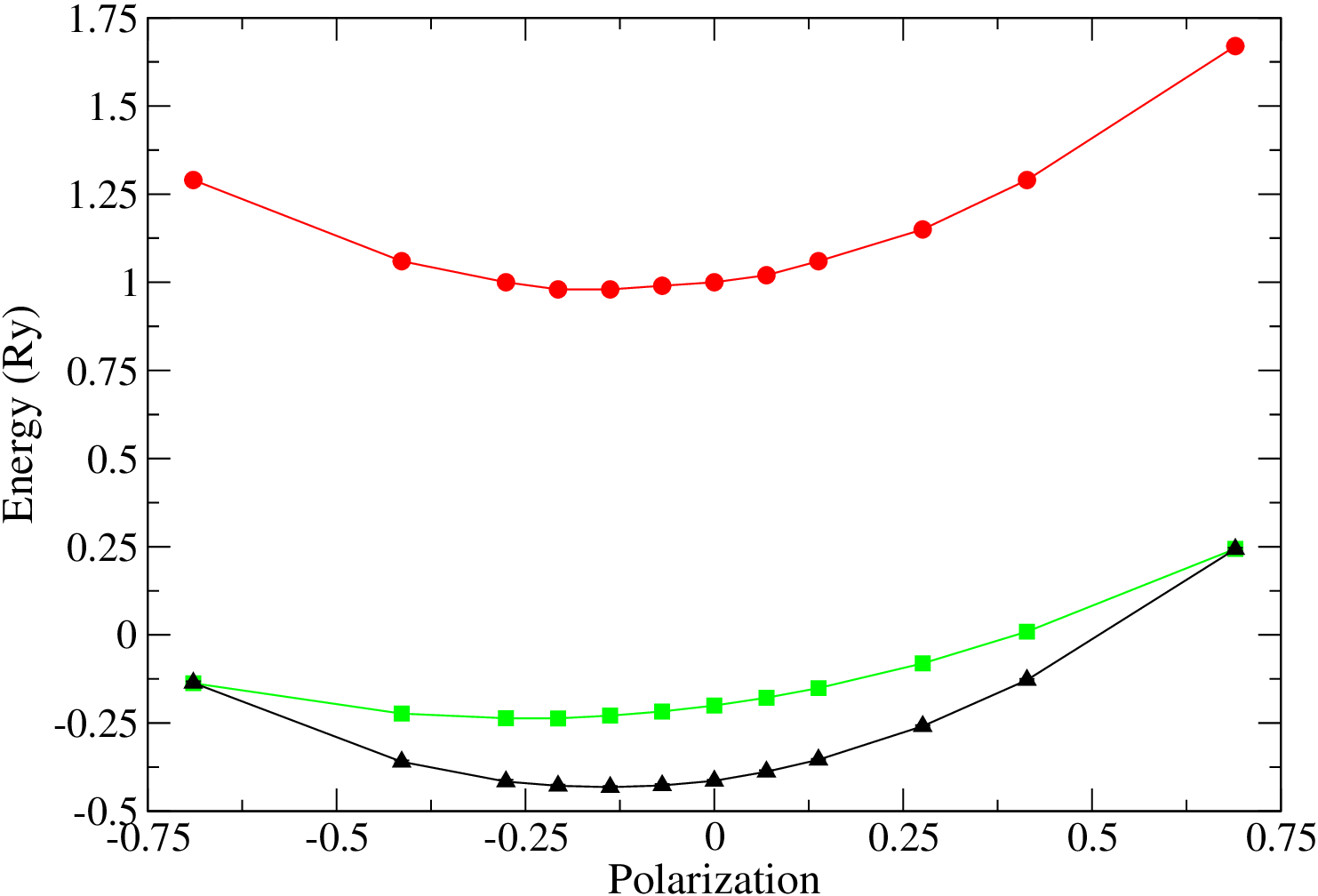}
\vspace{0.5cm}
\caption{(color online) Energy values at $r_s=1$ $\lambda =0.1$
respectively obtained with DMC (triangles), HF (squares) and without
Coulomb interaction (circles).} 
\label{rs1f}
\end{figure}

\begin{figure}[ht]
\vspace{0.7cm}
\centering
\includegraphics[scale=0.32]{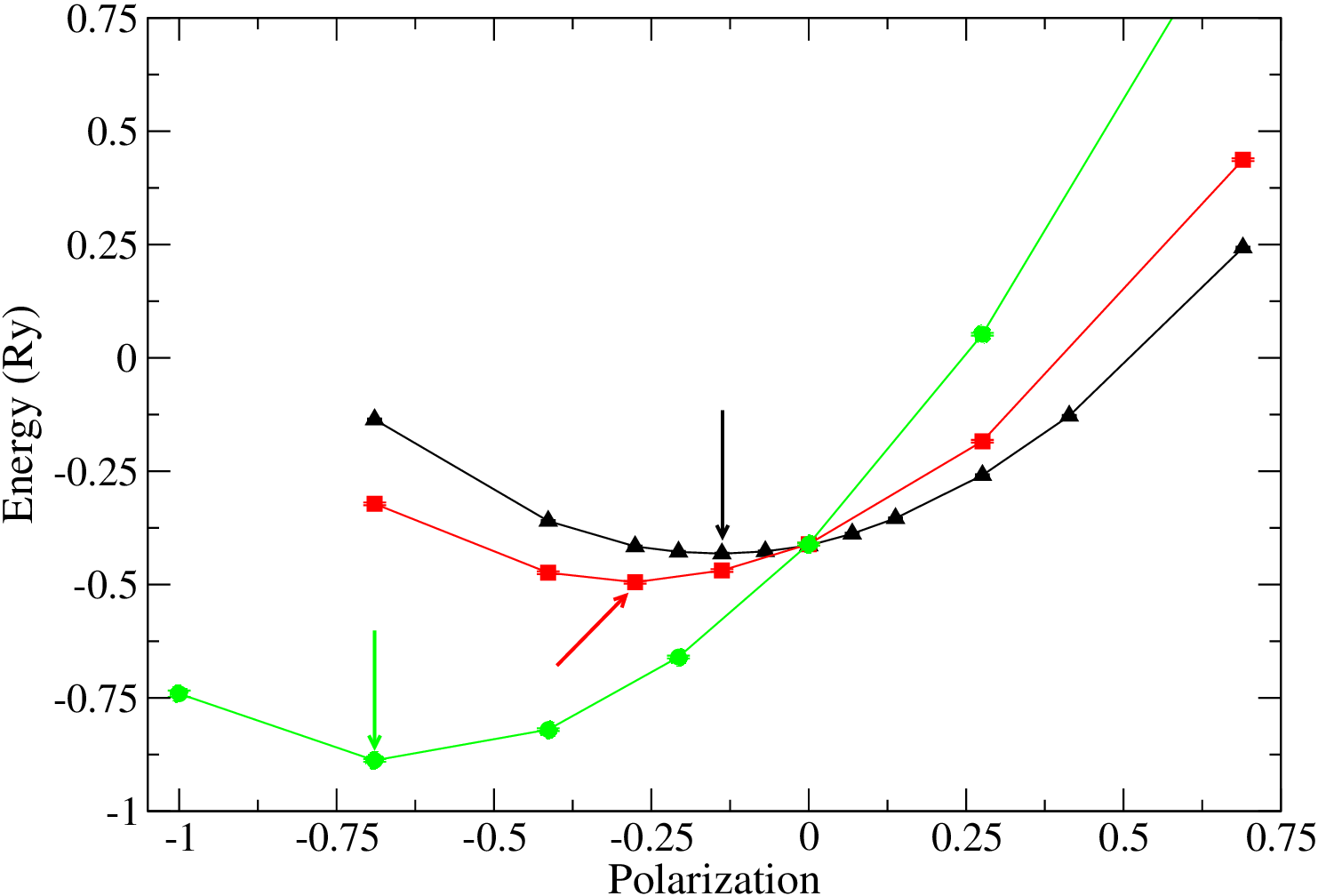}
\vspace{0.5cm}
\caption{(color online) Energy values obtained at $r_s=1$ for three
different $\lambda$ values, respectively $0.1$ (triangles), $0.2$
(squares) and $0.5$ (circles). 
The arrows indicate the point of minimum energy for each $\lambda$.
For increasing $\lambda$ the energy
minimum shifts to larger (in modulus) polarization.}
\label{triplof}
\end{figure}

DMC results at non zero polarization are obtained by projecting out of a
trial wave function $\psi_T$ whose Slater determinant contains different
numbers of $quasi-up$ and $quasi-down$ spin states. The approach is
apparently similar to what is commonly done for the electron gas in
absence of spin-orbit interaction. However, in this case the Hamiltonian does
not commute with the $z$ spin component. 
The choice of the quasi--up, quasi--down basis was made by comparison with
the results obtained using the standard spin--up and spin--down basis.
The latter consistently gives higher values of the energy for the ground 
state polarization.
Results in absence of Coulomb interaction predict spin
state polarization as a consequence of a two band dispersion law. A very
similar behavior for the energy as a function of polarization is found
in our results (see Fig. \ref{rs1f} ). This suggests the existence of
an analogous two band structure also in presence of Coulomb interaction
as experimentally proved by the Shubnikov-De Haas oscillations.

In Fig. \ref{triplof} energies at different $\lambda$ are shown as
a function of the polarization $\xi$. As expected, the polarization
depends on the intensity of the spin-orbit interaction, increasing in module with larger
$\lambda$. Calculations have been performed at different $r_s$ values,
corresponding to different densities. The onset of polarization occurs
at lower and lower strengths of the Rashba interaction when the density
is decreased. This is an effect of the Rashba interaction dependence
on the momentum.  In the Hamiltonian of Eq. \eqref{hamiltonian} the $1/r_s$
prefactors for the interaction terms, compared to the $1/r^2_s$ of the
kinetic energy let interactions become more efficient at high $r_s$
(low density) compared to the kinetic term. While at high density the
system will look less interacting, and therefore also less polarized,
at larger values of $r_s$ the system will be more sensitive to the the
spin-orbit interaction and will be more polarized also for smaller $\lambda$.

\begin{figure}[ht]
\vspace{0.7cm}
\centering
\includegraphics[scale=0.32]{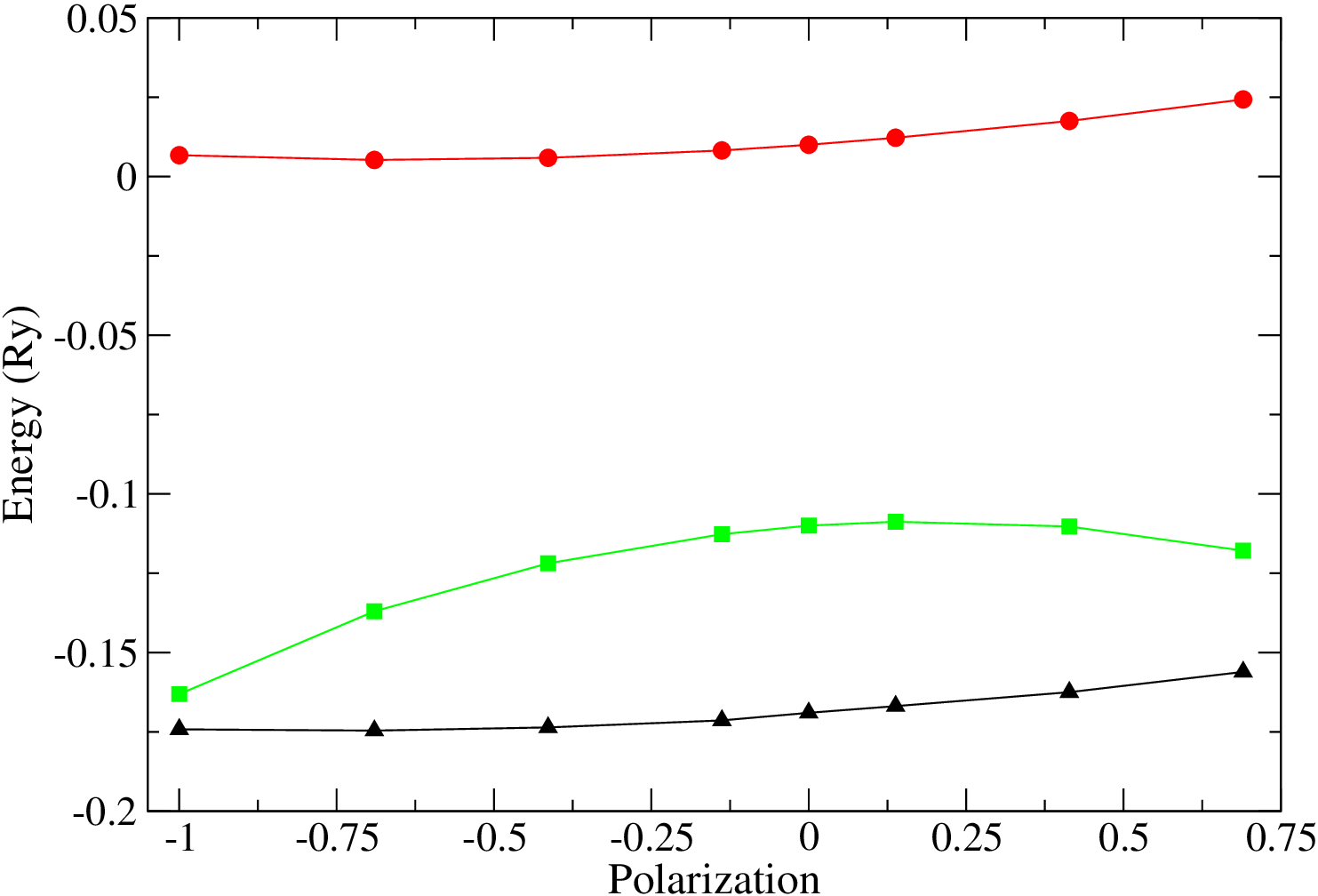}
\vspace{0.5cm}
\caption{(color online) Energy values at $r_s=10$ $\lambda =0.05$
as in the figure above. DMC results are represented by triangles, HF
by squares and results in absence of Coulomb interaction by circles.}
\label{rs10f}
\end{figure}

At lower densities correlations among electrons play a major role.
In Fig. \ref{rs1f} and \ref{rs10f} DMC energies are shown together with
the corresponding results from HF approximation. We see a worsening
agreement when $r_s$ increases. This must be due to the fact that HF
approximation does not take correlations into account, causing a larger
deviation when the N-electrons wave function becomes less similar to
a Slater determinant.

\begin{figure}[ht]
\vspace{0.7cm}
\centering
\includegraphics[scale=0.32] {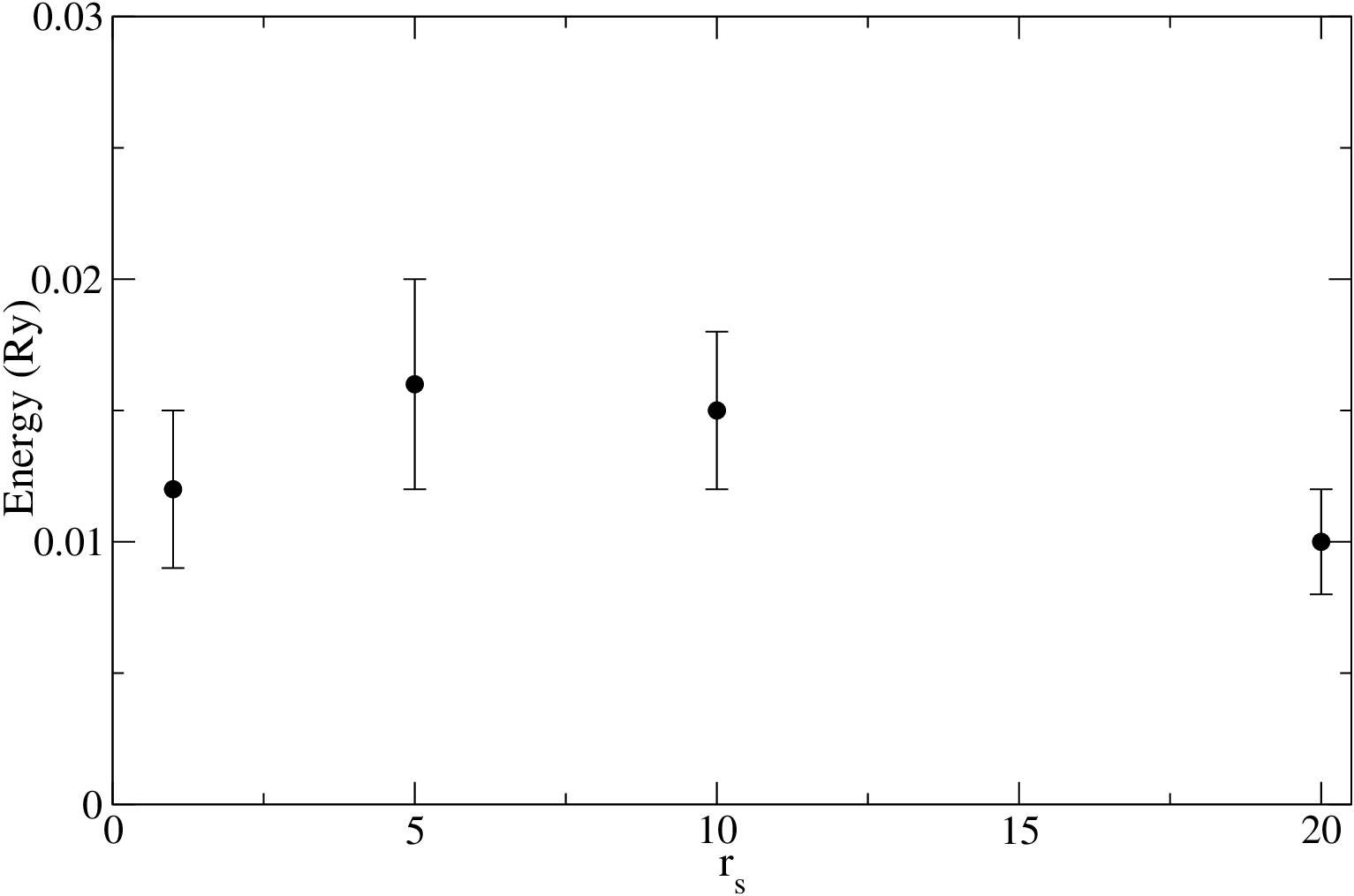}
\vspace{0.5cm}
\caption{Difference between energies without and in presence of Rashba
interaction ($\lambda=1$). Data were taken from \cite{attaccalite}.}
\label{cfrattaccaf}
\end{figure}

In Fig. \ref{cfrattaccaf} energies for
the 2D electron gas from Ref. \cite{attaccalite} are compared
with ground state results in presence of spin-orbit interaction with
fixed strength $\lambda=0.1$. In presence of Rashba interaction ground
states have a different structure and their polarization varies with
the density. However the ground state energy is always lower than in
absence of spin-orbit interaction over the whole density range.

\begin{figure}[ht]
\vspace{0.7cm}
\centering
\includegraphics[scale=0.32]{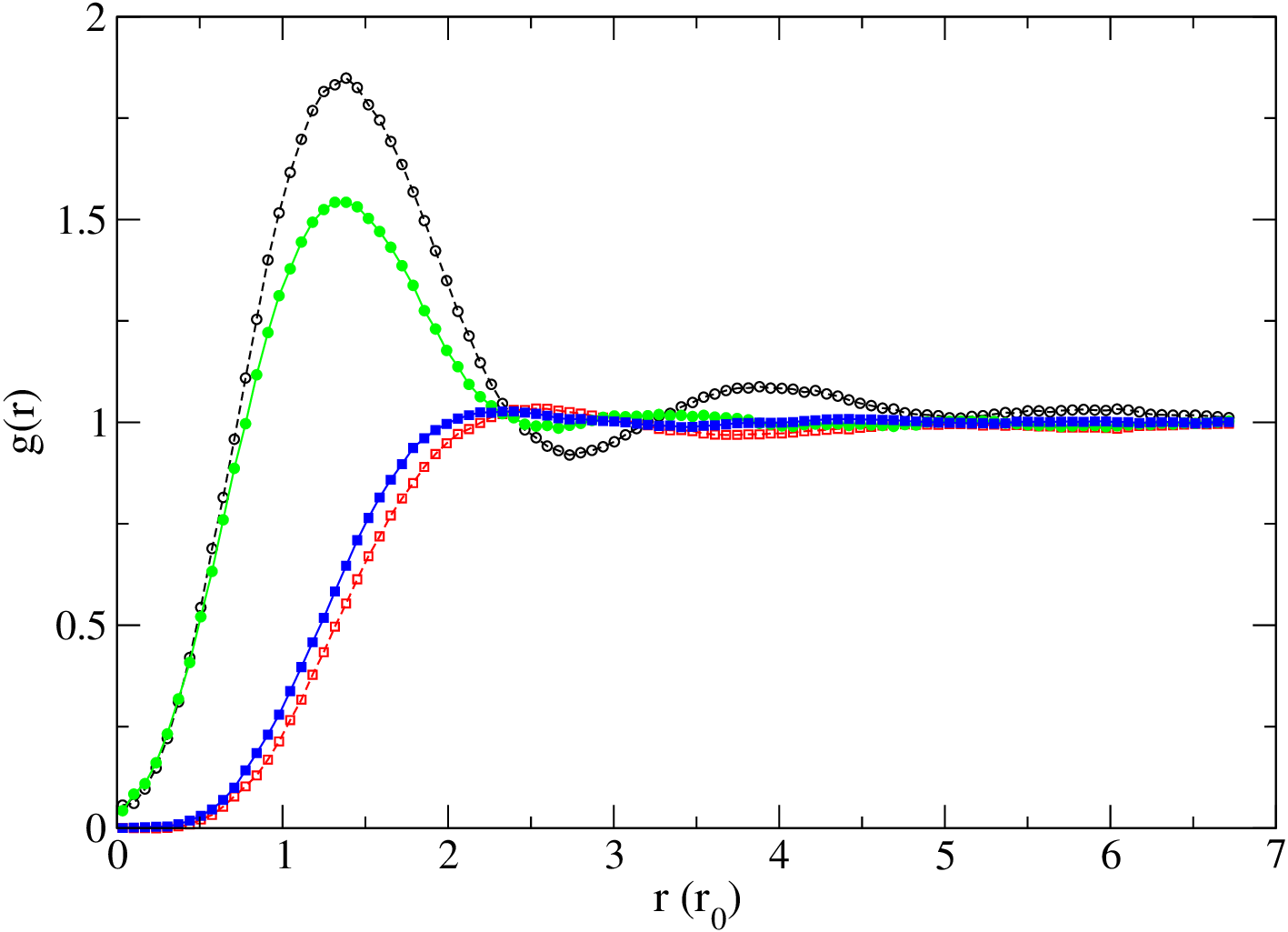}
\vspace{0.5cm}
\caption{(color online) Pair correlation function g(r) decomposed into 
singlet (circles) and triplet (squares) channels for $r_s=5$ with (closed 
symbols) and without (open symbols) spin-orbit interaction. 
For better comparison the curves are all normalized to give $g(r)\rightarrow 1$
for $r\rightarrow \infty$}
\label{gofrf} 
\end{figure}

In Fig. \ref{gofrf} the pair correlation function $g(r)$ decomposed into triplet
and singlet components is reported. Triplet and singlet components were
obtained collecting the following quantities: 
\begin{align} 
g_c (r)=N \sum_{i<j}\frac{\langle
\psi_T | \delta (r_{ij}-r) | \psi_T \rangle}{\langle \psi_T | \psi_T
\rangle}\\
g_{\sigma} (r)=N \sum_{i<j}\frac{\langle \psi_T | \delta
(r_{ij}-r) \boldsymbol\sigma_i\cdot \boldsymbol\sigma_j | \psi_T \rangle}{\langle
\psi_T | \psi_T \rangle}
\end{align}
where $N$ is a normalization factor, and composing them according 
to the triplet and singlet spin projectors\cite{gandolfi09}:
\begin{align}
g_{S=0}(r)=\frac{1}{4}[ g_c(r)-g_{\sigma}(r)]\\
g_{S=1}(r)=\frac{1}{4}[3 g_c(r)+g_{\sigma}(r)].
\end{align}
The results show a tendency
for the triplet and singlet first peaks to become smaller and closer to
eachother when spin-orbit interaction is switched on. This suggests a decrease
in the antiferromagnetic character of the electrons, probably due to
the induced spin polarization.

\section{Conclusions}
\label{conclusions}
We have implemented a DMC algorithm for treating Rashba spin-orbit
interaction in the electron gas. We have calculated
the equation of state of 2D electron gas in the presence of Rashba 
interaction. We computed the energy per particle as a function 
of the strength $\lambda$ of the Rashba potential, $r_s$ and the
polarization $\xi$ showing system polarization and comparing the results 
with analytical HF results. Our work not only gives results for the 2D
electron gas, but also provides a good test for the algorithm, opening
new possibilities for implementation on other systems, in particular
quantum dots or quantum wires involving both the Rashba and the 
Dresselhaus potential \cite{Dressel}.
A similar treatment can also be extended to 
finite systems such as atoms and molecules.

\section{acknowledgments}
This work was partially supported by INFM. Computer calculations were
performed on the WIGLAF cluster of the Department of Physics, University
of Trento, on the BEN cluster at ECT* in Trento, and also using CINECA
resources.


\appendix
\section{Tables with results}

In the following tables the Diffusion Monte Carlo results
are reported together with corresponding Hartree-Fock energies and results
for the system without the Coulomb interaction with the Rashba
interaction only.

\begin{table}[ht]
\caption{Energies of the system at $r_s=1$ and $\lambda=0.1$, $0.2$, and $0.5$
for various polarizations $\xi$ as defined in Eq. \eqref{eq:pol}.
The DMC result is computed with Coulomb and Rashba interaction, $E_{Rashba}$ is
the energy given by the Rashba Hamiltonian \eqref{hnorashba},
and in the last column the
Hartree--Fock result is reported.
All the energies are expressed in Ry.}
\begin{center}
\begin{tabular}{|ccc|ccc|}
\hline
\multicolumn{6}{|c|}{$\lambda=0.1$} \\
\hline 
$\xi$   & $N_-$ & $N_+$ & $E_{DMC}$ & $E_{Rashba}$ & $E_{HF}$ \\
\hline
0.690  & 9     & 49    &  0.243(4) & 1.67         & 0.244 \\
0.414  & 17    & 41    & -0.128(3) & 1.29         & 0.009 \\
0.276  & 21    & 37    & -0.259(3) & 1.15         & -0.081 \\
0.138  & 25    & 33    & -0.354(2) & 1.06         & -0.151 \\
0.069  & 27    & 31    & -0.388(4) & 1.024         & -0.178 \\
0.0     & 29    & 29    & -0.414(3) & 1.00         & -0.200 \\
-0.069     & 31    & 27    & -0.427(4) & 0.985      & -0.217 \\
-0.138   & 33    & 25    & -0.432(3) & 0.98         & -0.229 \\
-0.207   & 35    & 23    & -0.428(3) & 0.984        & -0.235 \\
-0.276   & 37    & 21    & -0.416(5) & 0.998        & -0.237 \\
-0.414   & 41    & 17    & -0.360(4) & 1.06         & -0.223 \\
-0.690   & 49    & 9     & -0.136(4) & 1.29         & -0.137 \\
\hline
\end{tabular}
\begin{tabular}{|ccc|ccc|}
\hline
\multicolumn{6}{|c|}{$\lambda=0.2$} \\
\hline 
$\xi$   & $N_-$ & $N_+$ & $E_{DMC}$ & $E_{Rashba}$ & $E_{HF}$ \\
\hline
0.690  & 9     & 49    &  0.437(4) & 1.86   &  0.435 \\
0.414  & 17     & 41    &  -0.008(4) &1.40  &  0.125 \\
0.276  & 21    & 37    & -0.184(5) & 1.23   & -0.022 \\
0.138  & 25    & 33    & -0.306(5) & 1.1    & -0.112\\
0.0     & 29    & 29    & -0.411(4) & 1.00  & -0.200 \\
-0.138   & 33    & 25    & -0.469(3) & 0.94 & -0.268 \\
-0.276   & 37    & 21    & -0.514(5) & 0.92 & -0.314\\
-0.414   & 41    & 17    & -0.474(5) & 0.94 & -0.340 \\
-0.690   & 49    & 9     & -0.322(4) & 1.09 & -0.328 \\
\hline
\end{tabular}
\begin{tabular}{|ccc|ccc|}
\hline
\multicolumn{6}{|c|}{$\lambda=0.5$} \\
\hline 
$\xi$   & $N_-$ & $N_+$ & $E_{DMC}$ & $E_{Rashba}$ & $E_{HF}$ \\
\hline
0.690  & 9     & 49    &  1.013(7) & 2.43 &  1.008 \\
0.276  & 21    & 37    &  0.053(6) & 1.47 &  0.230 \\
0.0     & 29    & 29    & -0.410(6) & 1.00 & -0.200 \\
-0.207   & 35    & 23    & -0.660(3) & 0.75 & -0.469 \\
-0.414   & 41    & 17    & -0.821(5) & 0.59 & -0.688 \\
-0.690   & 49    & 9     & -0.888(8) & 0.52 & -0.900 \\
-1.0     & 58    & 0     & -0.741(7) & 0.67 & -1.031 \\
\hline
\end{tabular}
\end{center}
\end{table}

\begin{table}[ht]
\begin{center}
\caption{Energies of the system at $r_s=5$ and $\lambda=0.02$, and $0.1$
for various polarizations $\xi$ as defined in Eq. \eqref{eq:pol}.
The DMC result is computed with Coulomb and Rashba interaction, $E_{Rashba}$ is
the energy given by the Rashba Hamiltonian \eqref{hnorashba}, and in the last column the
Hartree--Fock result is reported.
All the energies are expressed in Ry.}
\begin{tabular}{|ccc|ccc|}
\hline
\multicolumn{6}{|c|}{$\lambda=0.02$} \\
\hline 
$\xi$   & $N_-$ & $N_+$ & $E_{DMC}$ & $E_{Rashba}$ & $E_{HF}$ \\
\hline
0.690  & 9     & 49    & -0.2719(5) & 0.0667 & -0.2178 \\
0.414  & 17    & 41    & -0.2861(5) & 0.0515 & -0.2042 \\
0.138  & 25    & 33    & -0.2947(5) & 0.0423 & -0.1995 \\
0.0     & 29    & 29    & -0.2971(5) & 0.0400 & -0.2001 \\
-0.138   & 33    & 25    & -0.2978(4) & 0.0392 & -0.2026 \\
-0.414   & 41    & 17    & -0.2951(5) & 0.0422 & -0.2125 \\
-0.690   & 49    & 9     & -0.2867(4) & 0.0514 & -0.2330 \\
-1.0     & 58    & 0     & -0.2748(5) & 0.0693 & -0.2702 \\
\hline
\end{tabular}
\begin{tabular}{|ccc|ccc|}
\hline
\multicolumn{6}{|c|}{$\lambda=0.1$} \\
\hline 
$\xi$ & $N_-$ & $N_+$ & $E_{DMC}$ & $E_{Rashba}$ & $E_{HF}$ \\
\hline
0.690   & 9  & 49 & -0.2403(5) & 0.0972 & -0.1872 \\
0.414   & 17 & 41 & -0.2668(4) & 0.0701 & -0.1856 \\
0.138   & 25 & 33 & -0.2879(4) & 0.0486 & -0.1932 \\
0.0      & 29 & 29 & -0.2964(4) & 0.0400 & -0.2001 \\
-0.138    & 33 & 25 & -0.3033(3) & 0.0330 & -0.2088 \\
-0.414    & 41 & 17 & -0.3127(3) & 0.0236 & -0.2321 \\
-0.690    & 49 & 9  & -0.3155(5) & 0.0209 & -0.2635 \\
-1.0      & 58 & 0  & -0.3099(5) & 0.0267 & -0.3129 \\
\hline
\end{tabular}
\end{center}
\end{table}

\begin{table}[ht]
\begin{center}
\caption{Energies of the system at $r_s=10$ and $\lambda=0.02$, $0.05$ and $0.1$
for various polarizations $\xi$ as defined in Eq. \eqref{eq:pol}.
The DMC result is computed with Coulomb and Rashba interaction, $E_{Rashba}$ is
the energy given by the Rashba Hamiltonian \eqref{hnorashba}, and in the last column the
Hartree--Fock result is reported.
All the energies are expressed in Ry.}
\begin{tabular}{|ccc|ccc|}
\hline
\multicolumn{6}{|c|}{$\lambda=0.02$} \\
\hline 
$\xi$ & $N_-$ & $N_+$ & $E_{DMC}$ & $E_{Rashba}$ & $E_{HF}$ \\
\hline
0.690   &9  & 49 & -0.1621(5) & 0.0186 & -0.1236 \\
0.414   &17 & 41 & -0.1660(2) & 0.0140 & -0.1138 \\
0.138   &25 & 33 & -0.1688(3) & 0.0110 & -0.1099 \\
0.0      &29 & 29 & -0.1694(2) & 0.0100 & -0.1100 \\
-0.138    &33 & 25 & -0.1698(2) & 0.0094 & -0.1149 \\
-0.414    &41 & 17 & -0.1700(2) & 0.0094 & -0.1184 \\
-0.690    &49 & 9  & -0.1699(2) & 0.0109 & -0.1313 \\
-1.0      &58 & 0  & -0.1684(3) & 0.0147 & -0.1551 \\
\hline
\end{tabular}
\begin{tabular}{|ccc|ccc|}
\hline
\multicolumn{6}{|c|}{$\lambda=0.05$} \\
\hline 
$\xi$ & $N_-$ & $N_+$ & $E_{DMC}$ & $E_{Rashba}$ & $E_{HF}$ \\
\hline
0.690   &9  & 49 & -0.1561(4) & 0.0243 & -0.1179 \\
0.414   &17 & 41 & -0.1625(3) & 0.0175 & -0.1103 \\
0.138   &25 & 33 & -0.1669(3) & 0.0122 & -0.1088 \\
0.0      &29 & 29 & -0.1690(2) & 0.0100 & -0.1100 \\
-0.138    &33 & 25 & -0.1714(2) & 0.0082 & -0.1127 \\
-0.414    &41 & 17 & -0.1736(2) & 0.0059 & -0.1219 \\
-0.690    &49 & 9  & -0.1746(2) & 0.0052 & -0.1370 \\
-1.0      &58 & 0  & -0.1742(2) & 0.0067 & -0.1631 \\
\hline
\end{tabular}
\begin{tabular}{|ccc|ccc|}
\hline
\multicolumn{6}{|c|}{$\lambda=0.1$} \\
\hline 
$\xi$ & $N_-$ & $N_+$ & $E_{DMC}$ & $E_{Rashba}$ & $E_{HF}$ \\
\hline
0.690   &9  & 49 & -0.1459(4) & 0.0338 & -0.1084 \\
0.414   &17 & 41 & -0.1565(3) & 0.0233 & -0.1045 \\
0.138   &25 & 33 & -0.1656(3) & 0.0141 & -0.1068 \\
0.0      &29 & 29 & -0.1696(2) & 0.0100 & -0.1100 \\
-0.138    &33 & 25 & -0.1733(2) & 0.0063 & -0.1146 \\
-0.414    &41 & 17 & -0.1792(2) & 0.0001 & -0.1277 \\
-0.690    &49 & 9  & -0.1837(2) & -0.0043 & -0.1465 \\
-1.0      &58 & 0  & -0.1861(2) & -0.0067 & -0.1764 \\
\hline
\end{tabular}
\end{center}
\end{table}

\begin{table}[ht]
\caption{Energies of the system at $r_s=20$ and $\lambda=0.01$, $0.02$ and $0.1$
for various polarizations $\xi$ as defined in Eq. \eqref{eq:pol}.
The DMC result is computed with Coulomb and Rashba interaction, $E_{Rashba}$ is
the energy given by the Rashba Hamiltonian \eqref{hnorashba}, and in the last column the
Hartree--Fock result is reported.
All the energies are expressed in Ry.}
\begin{tabular}{|ccc|ccc|}
\hline
\multicolumn{6}{|c|}{$\lambda=0.01$} \\
\hline 
$\xi$ & $N_-$ & $N_+$ & $E_{DMC}$ & $E_{Rashba}$ & $E_{HF}$ \\
\hline
0.690 &  9  & 49 & -0.0901(3) & 0.0046 & -0.0665 \\
0.414 &  17 & 41 & -0.0912(2) & 0.0035 & -0.0604 \\
0.138 &  25 & 33 & -0.0919(2) & 0.0027 & -0.0577 \\
0.0    &  29 & 29 & -0.0921(2) & 0.0025 & -0.0575 \\
-0.138  &  33 & 25 & -0.0923(2) & 0.0024 & -0.0581 \\
-0.414  &  41 & 17 & -0.0923(3) & 0.0023 & -0.0616 \\
-0.690  &  49 & 9  & -0.0920(3) & 0.0017 & -0.0684 \\
-1.0    &  58 & 0  & -0.0912(4) & 0.0037 & -0.0812 \\
\hline
\end{tabular}
\begin{tabular}{|ccc|ccc|}
\hline
\multicolumn{6}{|c|}{$\lambda=0.02$} \\
\hline 
$\xi$ & $N_-$ & $N_+$ & $E_{DMC}$ & $E_{Rashba}$ & $E_{HF}$ \\
\hline
0.690 & 9  & 49 & -0.0892(2) & 0.0056 & -0.0655 \\
0.414 & 17 & 41 & -0.0906(2) & 0.0041 & -0.0598 \\
0.138 & 25 & 33 & -0.0918(3) & 0.0029 & -0.0575 \\
0.0    & 29 & 29 & -0.0922(2) & 0.0025 & -0.0575 \\
-0.138  & 33 & 25 & -0.0924(2) & 0.0022 & -0.0583 \\
-0.414  & 41 & 17 & -0.0928(2) & 0.0018 & -0.0622 \\
-0.690  & 49 & 9  & -0.0928(3) & 0.0018 & -0.0693 \\
-1.0    & 58 & 0  & -0.0924(5) & 0.0023 & -0.0825 \\
\hline
\end{tabular}
\begin{tabular}{|ccc|ccc|}
\hline
\multicolumn{6}{|c|}{$\lambda=0.1$} \\
\hline 
$\xi$ & $N_-$ & $N_+$ & $E_{DMC}$ & $E_{Rashba}$ & $E_{HF}$ \\
\hline
0.690  & 9  & 49 & -0.0811(3) & 0.0132 & -0.0579 \\
0.414  & 17 & 41 & -0.0856(2) & 0.0087 & -0.0552 \\
0.0     & 29 & 29 & -0.0918(2) & 0.0045 & -0.0560 \\
-0.138   & 33 & 25 & -0.0937(2) & 0.0006 & -0.0599 \\
-0.414   & 41 & 17 & -0.0972(2) & -0.0029 & -0.0668 \\
-0.690   & 49 & 9  & -0.1001(2) & -0.0058 & -0.0770 \\
-1.0     & 58 & 0  & -0.1026(2) & -0.0083 & -0.0932 \\
\hline
\end{tabular}
\end{table}

\end{document}